\documentclass[a4paper]{jpconf}
\usepackage{graphicx}
\usepackage{iopams}
\usepackage{hyperref}
\usepackage{multirow}

\newcommand{\xdownarrow}[1]{%
  {\left\downarrow\vbox to #1{}\right.\kern-\nulldelimiterspace}
}

\begin{document}
\title{Federating distributed storage for clouds in ATLAS}

\author{F~Berghaus$^1$, K~Casteels$^1$, A~Di Girolamo$^2$, C~Driemel$^1$, M~Ebert$^1$, F~Furano$^2$, F~Galindo$^3$, M~Lassnig$^2$, C~Leavett-Brown$^1$, M~Paterson$^1$, C~Serfon$^2$, R~Seuster$^1$, R~Sobie$^1$, R~Tafirout$^3$ and R~P~Taylor$^1$}
\address{$^1$ Department of Physics and Astronomy, University of Victoria, Finnerty Road, Victoria V8P~5C2, Canada}
\address{$^2$ CERN, Geneva 1211, Switzerland}
\address{$^3$ TRIUMF, Wesbrook Mall, Vancouver V6T~2A3 Canada}

\ead{frank.berghaus@cern.ch}

\begin{abstract}
Input data for applications that run in cloud computing centres can be stored at distant repositories, often with multiple copies of the popular data stored at many sites. Locating and retrieving the remote data can be challenging, and we believe that federating the storage can address this problem. A federation would locate the closest copy of the data on the basis of GeoIP information. Currently we are using the dynamic data federation Dynafed, a software solution developed by CERN IT. Dynafed supports several industry standards for connection protocols like Amazon's S3, Microsoft's Azure, as well as WebDAV and HTTP. Dynafed functions as an abstraction layer under which protocol-dependent authentication details are  hidden from the user, requiring the user to only provide an X509 certificate. We have setup an instance of Dynafed and integrated it into the ATLAS data distribution management system. We report on the challenges faced during the installation and integration. We have tested ATLAS analysis jobs submitted by the PanDA production system and we report on our first experiences with its operation.
\end{abstract}

\section{Introduction}
We aim to run data-intensive applications on globally distributed opportunistic resources that have no local grid storage. The ATLAS~\cite{atlas} experiment leverages a globally distributed system of infrastructure as a service (IaaS) clouds as part of its distributed computing system. These resources are integrated into the ATLAS distributed computing system using the Cloud Scheduler~\cite{cloud-scheduler} technology developed at the University of Victoria. These IaaS resources are used opportunistically, and do not support any local grid infrastructure.

The workflows executed by high energy physics experiments often demand large volumes of input data or produce a significant volume of output data. We aim to use a data federation, such as Dynafed~\cite{dynafed}, to redirect the applications running on opportunistic resources to the optimal storage endpoint to retrieve input or deposit output data. We also aim to integrate storage solutions offered by cloud providers into the ATLAS distributed data management system using Dynafed.

In this paper we explain a system leveraging Cloud Scheduler and Dynafed which successfully executed functional test jobs as part of the ATLAS distributed computing system on the CERN OpenStack~\cite{openstack} cloud resource that read their input from and wrote their output to an object store implemented using Ceph~\cite{ceph} and exposing an S3 compatible gateway.

\section{Conceptual Design}
\begin{figure}
  \centering
  \includegraphics[width=\textwidth]{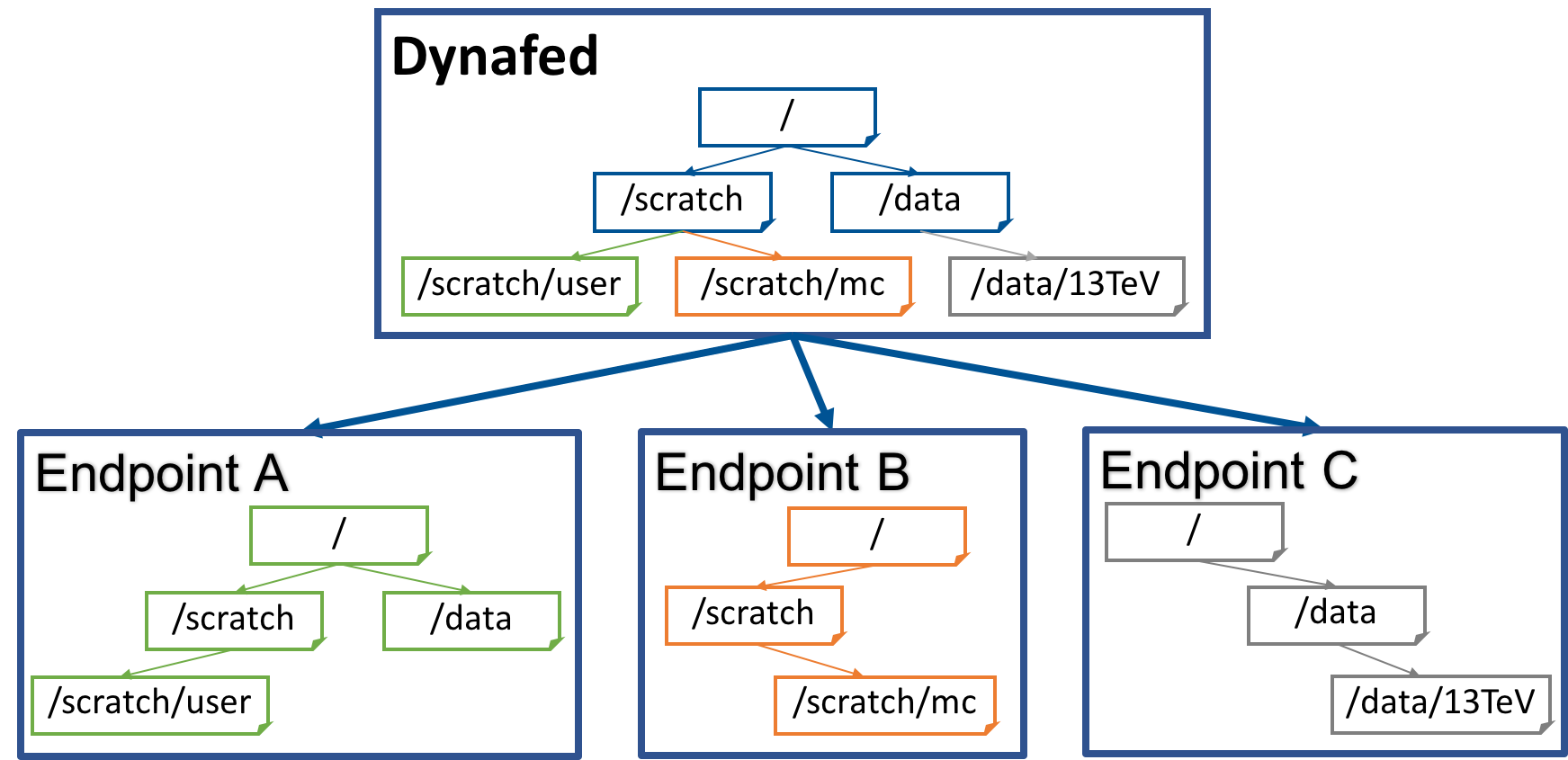}
  \caption{The dynamic federation is connected to multiple endpoints. Each endpoint may be a file system or an object store accessible using a protocol which allows redirection, such as HTTP. The dynamic federation provides a namespace that is a union of all the namespaces of the endpoints. That namespace is presented as a regular directory structure by the Dynafed. The content of a displayed directory is calculated when accessed.}
  \label{fig:conceptual-design}
\end{figure}

The ATLAS experiment leverages the resources of the Worldwide LHC Computing Grid, WLCG~\cite{wlcg}. The computing centres that are part of the WLCG and support ATLAS provide a global storage infrastructure for the experiment data and simulated events.
%While the central ATLAS computing infrastructure uses purpose-specific protocols to access the content of these grid storage elements,
They may be accessed using standard protocols, such as HTTP with WebDAV extensions. Dynafed supports storage backends that offer HTTP and WebDAV access and promises sufficient scalability to create the appearance of a single virtual namespace for the entire ATLAS data catalogue.  Figure~\ref{fig:conceptual-design} shows how Dynafed could unify the namespaces of attached storage elements into a single namespace.

Dynafed allows the usage of cloud storage systems such as S3, SWIFT, and Azure. On the user-facing side Dynafed still presents an HTTP interface implementing authentication and authorization through an X509 public key infrastructure. Dynafed supports grid security infrastructure extensions of X509 with VOMS attributes~\cite{voms}. Credentials may be presented as certificate and key or as a proxy, which allows the additional use of VOMS attributes. When Dynafed forwards clients to cloud storage systems, it translates their X509 credentials to pre-signed URL that permit, for a limited time, access to the cloud storage system.

\section{Data Access}
The dynamic federation used for this work was configured to use three endpoints: one at CERN, one at TRIUMF, and one at the University of Victoria. Each endpoint was a CephS3 object store. Table~\ref{tab:dynafed-arch} illustrates the task division within the dynamic federation to handle client requests.

To access data through Dynafed a client makes a request for a file using HTTP optionally with WebDAV extensions.
%WebDAV is the appropriate protocol for interacting with files and file systems, so
We will focus on WebDAV from here on. In our configuration the Dynafed only allows access to members of the ATLAS Virtual Organization. The client must provide X509 credentials with the request. The credential must be signed by a trusted certificate authority. If the credential contains VOMS extensions certifying the user to be a member of the ATLAS collaboration, access is granted. Without VOMS extensions the Dynafed checks the credential against all current members of the ATLAS collaboration and grants access if a match is found. Administrators and data management services have privileged accounts. Authorized clients are redirected to a signed URL on the closest CephS3 endpoint. Authorization is granted explicitly for reading, writing, listing, and/or deleting operations.

When a file is requested, the dynamic federation checks whether the locations of the file are already in its cache. If so, the cached entry is used, otherwise each endpoint is queried for the file after name translation to that endpoint. Dynafed waits for responses\footnote{Up to a given timeout set to 3~seconds here} to collect and cache. The resulting endpoints are evaluated for proximity to the client and the client is redirected to the closest copy. The Dynafed regularly polls all connected endpoints to determine if they are reachable. Should an endpoint be unresponsive, requests will not be forwarded to it.
% This should increase the stability of the storage system as a whole by dynamically adapting to storage endpoint failures.

\begin{table}
  \centering
  \caption{The dynamic web federation is an Apache server running the LCGDM implementation of WebDAV. The namespace usually managed by LCGDM has been replaced by the uniform general redirector (UGR) which translates the requests to the web file system to the connected endpoints. The endpoint modules handle the communication with the configured endpoints. All requests are cached in memory on the server as well as in a second-level cache which may be shared across multiple load-balanced servers.}
  \label{tab:dynafed-arch}
  \begin{tabular}{ lllp{10cm} }
  \br
  \multicolumn{3}{ l }{Component} & Purpose \\
  \mr
  \multirow{8}{*}{$\xdownarrow{1.8cm}$} & \multicolumn{2}{ l }{Apache} & Load the \texttt{lcgdm\_dav} module and start up a WebDAV server \\ \cline{2-4}
   & \multicolumn{2}{ l }{\texttt{lcgdm\_dav}} & \multirow{2}{10cm}{Configure dmlite and load the uniform general redirector as namespace plugin} \\
   & \multicolumn{2}{ l }{dmlite} & \\ \cline{2-4}
   & \multicolumn{2}{ l }{UGR} & Configure authentication and endpoints \\ \cline{3-4}
   & \multirow{4}{*}{\rotatebox{90}{Plugins}} & dmlite & \multirow{4}{10cm}{Communicate with endpoints on request} \\
   &  & WebDAV/HTTP &  \\
   &  & S3 &  \\
   &  & Azure &  \\ \hline
   & \multicolumn{2}{ l }{Memcached} & Cache recent redirects to distributed object caching system \\
  \br
  \end{tabular}
\end{table}

 %It is also possible to query a metalink which returns a XML list of all copies of the queried file. In the future we wish to implement chunked downloads from multiple locations using this metalink and the aria2c copy tool. Should the client make a request to write, Dynafed redirects the client to the geographically closest writeable storage element.

\section{Application Workflow}
In order to integrate the dynamic federation into the ATLAS distributed computing and data management system, it was defined as a storage element associated with the \texttt{CERN-EXTENSION}\footnote{\texttt{CERN-EXTENSION} is an ATLAS site defined as a part of the \texttt{CERN-PROD} WLCG and GOCDB site.} ATLAS site. It was configured to be accessible using WebDAV and flagged as special in the ATLAS grid information system to allow Rucio~\cite{rucio} to select a copy tool implementation which does not move or rename files.

The input datasets for analysis and production functional tests were transferred to the dynamic federation using the file transfer service at CERN. Once the transfers completed successfully the data was registered manually in the Rucio data catalogue.

\begin{figure}
  \includegraphics[width=\textwidth]{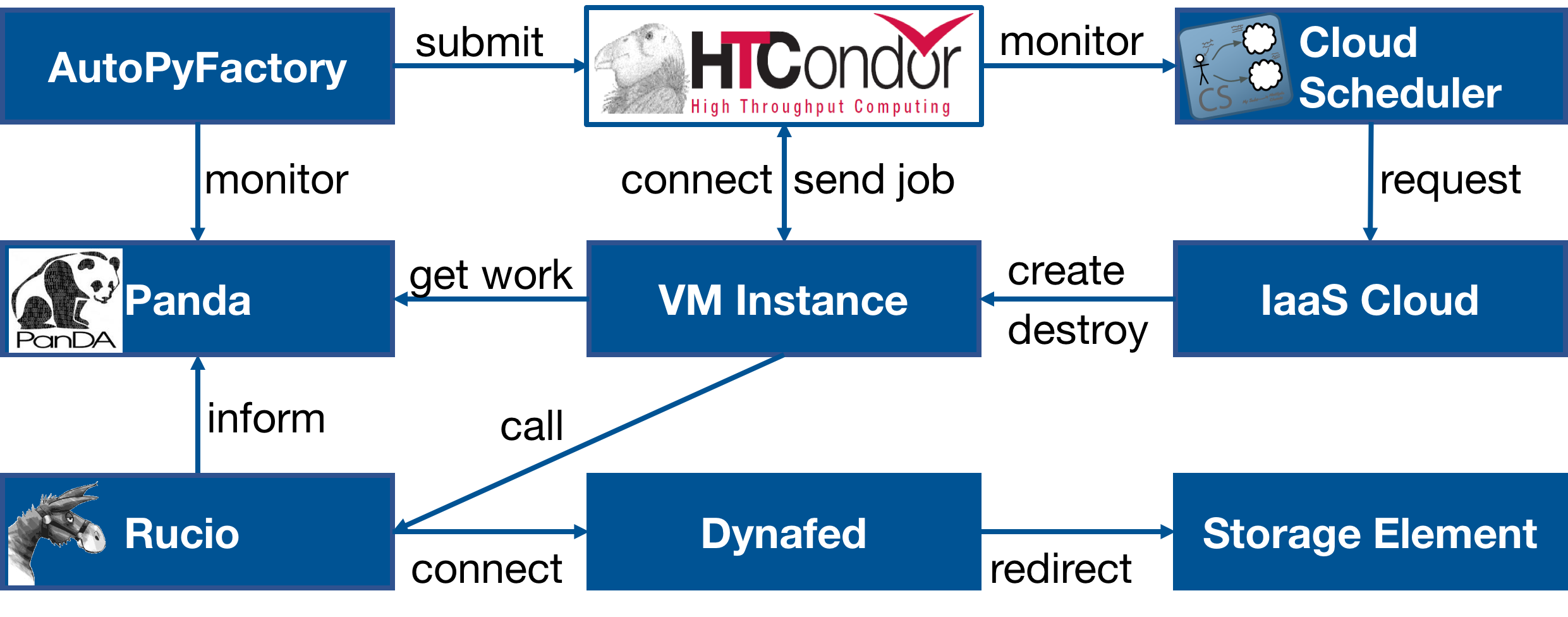}
  \caption{A client (AutoPyFactory~\cite{apf} or Harvester~\cite{harvester}) submits pilot wrapper scripts to an HTCondor~\cite{htcondor} queue. The queue is monitored by a Cloud Scheduler. The Cloud Scheduler makes requests to connected cloud interfaces in response to queued jobs. The cloud infrastructures create virtual machine instances and provide user-data to cloud-init running in the virtual machines for configuration. The VM instances are configured to connect to HTCondor and start consuming jobs from the queue. The jobs are pilot wrapper scripts which download the pilot. The pilot gets tasks from a PanDA~\cite{panda} queue and uses Rucio to download input data and upload results. Rucio is configured to contact the dynamic federation using the HTTP protocol with WebDAV extensions. The federation forwards the Rucio client to the closest available storage element.}
  \label{fig:atlas-cloud}
\end{figure}

With the input data registered in the data catalogue it was possible to run grid jobs against the data in the dynamic federation. The jobs were executed on virtual machines hosted on the CERN OpenStack using the Cloud Scheduler technology as illustrated in Figure~\ref{fig:atlas-cloud}. The resulting data and logs were uploaded to the CephS3 storage via Dynafed upon job completion.

Some additional development is required for full integration of cloud storage into the ATLAS distributed data management: bulk transfers negotiated between storage endpoints using the HTTP protocol must be fully supported, and the data management system must be able to parse the checksums of files on cloud storage.
% The first point requires upgrades to the logic of Rucio's conveyer mechanism. The second point requires Rucio to fully support MD5~\cite{md5} checksums, which are commonly used in cloud storage systems, along side ADLER32~\cite{adler32} checksums which are the standard on the worldwide LHC computing grid. This will likely involve some additional logic to the grid file access libraries since these checksums are also communicated in a different fashion on cloud storage as opposed to grid storage~\cite{content-md5, request-digest}.

\section{Summary}
It was shown that ATLAS jobs can retrieve and deposit their data on a cloud storage system accessed via a dynamic federation using the HTTP protocol with WebDAV extensions. The jobs ran on virtual machine instances in a cloud and could be scheduled anywhere in the distributed cloud system currently running as part of the ATLAS production system. Further development is necessary to allow the execution of production or analysis jobs against the dynamic federation. Work is ongoing to integrate the dynamic federation with the Belle~II~\cite{belle2} experiment and the DIRAC~\cite{dirac} workload management system. While this development is being pursued against opportunistic cloud resources it should also be useful in the context of volunteer resources~\cite{boinc}.

\end{document}